\newif\ifauthor\authortrue{}

\ifauthor{}
\documentclass[sigplan,screen,dvipsnames,authorversion]{acmart}
\else
\documentclass[sigplan,screen,dvipsnames]{acmart}
\fi

\usepackage[utf8]{inputenc}
\usepackage[T1]{fontenc}

\usepackage{microtype}
\usepackage{prettyref}
\usepackage{xcolor}
\usepackage{xspace}
\usepackage{enumitem}
\usepackage{subcaption}
\usepackage{booktabs}
\usepackage{listings}

\definecolor{darkblue}{rgb}{0,0,0.5}
\definecolor{darkgreen}{rgb}{0,0.3,0}
\definecolor{darkpink}{rgb}{0.4,0,0.3}
\definecolor{graygreen}{rgb}{0.3,0.5,0.3}
\definecolor{grayblue}{rgb}{0.2,0.2,0.6}
\definecolor{grayred}{rgb}{0.5,0.2,0.2}

\newcommand*{\basecodestyle}{\ttfamily}
\newcommand*{\codestyle}{\small\basecodestyle}
\newcommand*{\footnotecodestyle}{\footnotesize\basecodestyle}
\newcommand*{\kwstyle}{\bf\codestyle\color{Blue}}

\newcommand\cd[1]{\lstinline[breakatwhitespace]{#1}}
\newcommand\cdfoot[1]{\lstinline[breakatwhitespace,basicstyle=\footnotecodestyle]{#1}}

\newcommand{\pref}[1]{\prettyref{#1}}

\newrefformat{fig}{Figure~\ref{#1}}
\newrefformat{aside}{Aside~\ref{#1}}

\newcommand{\rebound}[0]{\textsc{Rebound}\xspace}
\newcommand{\piforall}[0]{\texttt{pi-forall}\xspace}
\newcommand\Bound[0]{\texttt{Bound}\xspace}
\newcommand\unbound[0]{\texttt{unbound}\xspace}

\setcopyright{acmlicensed}
\acmDOI{10.1145/3759164.3759348}
\acmYear{2025}
\copyrightyear{2025}
\acmISBN{979-8-4007-2147-2/25/10}
\acmConference[Haskell '25]{Proceedings of the 18th ACM SIGPLAN International Haskell Symposium}{October 12--18, 2025}{Singapore, Singapore}
\acmBooktitle{Proceedings of the 18th ACM SIGPLAN International Haskell Symposium (Haskell '25), October 12--18, 2025, Singapore, Singapore}
\received{2025-06-09}
\received[accepted]{2025-07-17}

\begin{document}
\title{Rebound: Efficient, Expressive, and Well-Scoped Binding}

\author{Noé De Santo}
\orcid{0009-0006-5119-3895}
\affiliation{%
  \institution{University of Pennsylvania}
  \city{Philadelphia}
  \country{USA}
}
\email{ndesanto@seas.upenn.edu}

\author{Stephanie Weirich}
\orcid{0000-0002-6756-9168}
\affiliation{%
  \institution{University of Pennsylvania}
  \city{Philadelphia}
  \country{USA}
}
\email{sweirich@seas.upenn.edu}

\begin{abstract}
  We introduce the \rebound{} library that supports well-scoped term
  representations in Haskell and automates the definition of substitution,
  alpha-equivalence, and other operations that work with binding structures. The
  key idea of our design is the use of first-class environments that map
  variables to expressions in some new scope.  By statically tracking scopes,
  users of this library gain confidence that they have correctly maintained the
  subtle invariants that stem from using de Bruijn indices. Behind the scenes,
  \rebound{} uses environments to optimize the application of substitutions,
  while providing explicit access to these data structures when desired. We
  demonstrate that this library is expressive by using it to implement a wide
  range of language features with sophisticated uses of binding and several
  different operations that use this abstract syntax.  Our examples include
  \piforall, a tutorial implementation of a type checker for a dependently-typed
  programming language. Finally, we benchmark \rebound{} to understand its
  performance characteristics and find that it produces faster code than
  competing libraries.
\end{abstract}



\maketitle

\section{Implementing Binding}

Implementors of programming languages, logics and calculi in Haskell have a
choice to make when it comes to representing the binding structure of their
programming languages. They need a representation of variables and binding
locations (such as $\lambda$-expressions) that accurately represents their
abstract syntax and operations that use this syntax (such as substitution,
evaluation, and type checking). There are many binding representations
possible~\cite{debruijn,washburn:bgb-journal,phoas,mcbride:locally-nameless}
and in this choice, implementors must balance multiple factors. In general,
they would like one that is simple to work with, requires minimal boilerplate,
and gives them confidence that their code is correct. At the same time, they
would like an efficient implementation that does not slow down their code.

In the setting of mechanized programming language semantics, it is common to
use de Bruijn indices with a \emph{scope-safe}
representation~\cite{altenkirch-reus,debruijn-coq,schaefer:autosubst,stark:autosubst2,allais:kit}. In
this case, the abstract syntax tree uses a dependent-type index to statically
track the number of free variables currently in scope. The types of operations
that work with syntax describe how they modify the current scope, and the
type-checker statically verifies the correctness of this specification.

Haskell programmers have seen scope-safe representations of lambda calculus
terms before, most notably in a functional pearl by Bird and
Paterson~\cite{bird:debruijn}, which uses a nested datatype to statically
track scoping level, and in the \Bound{} library~\cite{kmett:bound}, which
optimizes this representation using an explicit weakening operation. However,
despite the long history, this approach is not widely used in practice.

Therefore, we have developed the \emph{\rebound{}} library\footnote{Available at
\url{https://github.com/sweirich/rebound}.} as a tool to assist Haskell
developers in working with well-scoped de Bruijn indices. This tool provides
type classes, abstract data types, and can automatically derive necessary
operations for working with variables. It is also accompanied by a suite of
literate examples that demonstrate its use in various settings.

The design of this library is governed by three goals:

\begin{description}[leftmargin=0pt,listparindent=0pt]
\item[Correctness] \rebound{} uses Dependent Haskell to statically track
  the scopes of bound variables. Because variables are represented by de
  Bruijn indices, scopes are represented by natural numbers, bounding the
  indices that can be used. If the scope is 0, then the term must be
  closed. The type checker can identify when users violate the subtle
  invariants of working with indices.

\item[Efficiency] Behind the scenes, \rebound{} uses first-class parallel
  substitution, or \emph{environments}, to delay the execution of operations
  such as shifting and substitution. Furthermore, these environments are
  accessible to library users who would like fine control over when these
  operations happen.

\item[Convenience] \rebound{} is based on a type-directed approach to binding,
  where users indicate binding structure in their abstract syntax through the
  use of types provided by the library.  As a result, \rebound{} provides a
  clean, uniform, and automatic interface to common operations such as
  substitution, alpha-equivalence, and free variable calculation.
\end{description}

Our goal with this paper is to highlight the key ideas that underlie the
design of this library, to describe the design space and potential trade-offs
in its implementation, and to evaluate its usability and performance at scale.

In Section~\ref{sec:examples}, we develop the key idea that underlies our
approach: the use of parallel substitutions, which we call
\emph{environments}~\cite{abadi:explicit-substitutions}.  A well-scoped
environment is a finite map from variable indices, bounded by some natural number $n$, to
expressions with indices bounded by $m$. These two numbers are part of
the environment's static type and ensure that we only ever look up indices that
are valid in the current scope and that we know the scoping of the resulting
term after the substitution has been applied. Section~\ref{sec:examples} demonstrates that, by
considering how substitutions may be composed and delayed during the execution
of an evaluator for the untyped lambda calculus, we are able to dramatically
improve runtime performance.

To make this idea readily available to Haskell programmers, in
Section~\ref{sec:library}, we show that the key ideas can be packaged up
inside appropriate type classes and abstract types, providing novice users
with a simple interface to these operations, optionally supported by generic programming
to eliminate boilerplate~\cite{magalhaes:generics}.  To evaluate the expressiveness
of the \rebound{} library and provide a tutorial on its usage, we have
collected a suite of examples that challenge the capabilities of the
library. We give an overview of this suite in Section~\ref{sec:patterns} and
demonstrate its support for various forms of \emph{pattern binding}.

Furthermore, we have developed a case study to evaluate this work in the context
of a practical setting. Section~\ref{sec:pi-forall} discusses our adaptation of
the implementation of the \piforall{} language~\cite{weirich:pi-forall} to use
this library. This code base includes a parser, scope checker and bidirectional
type checker for a dependently-typed programming language with datatypes and
dependent pattern matching.

When developing the library, we have choices about how we may \emph{represent}
environments in Haskell and with how functions that operate over lambda calculus
terms may \emph{use} environments as part of their operation. In
Section~\ref{sec:benchmarks}, we deepen our performance analysis by benchmarking
various environment representations. Furthermore, to understand how our use of
de Bruijn indices compares against other approaches, we benchmark uses of
\rebound{} against other binding libraries available in Haskell. We also have
benchmarked the performance of \piforall{} using \rebound{} against its prior
implementation. In each case we find that \rebound{} outperforms its
competition, especially on benchmarks that require repeated $\beta$-reductions.
\footnote{Our benchmarking code is also available at
\url{https://github.com/sweirich/rebound}.}

The use of a well-scoped representation is a form of dependently-typed
programming in Haskell. While our examples and case studies provide positive
evidence that this approach is beneficial, we acknowledge that there are
trade-offs. We identify limitations with working with a scope-safe
representation in Section~\ref{sec:discussion}.  Finally, we discuss related
work in Section~\ref{sec:related-work} and conclude in
Section~\ref{sec:conclusion}.

\section{Well-Scoped Interpreters}%
\label{sec:examples}

Consider the following \emph{scope-safe} representation of the lambda
calculus:
\begin{lstlisting}
data Nat = Z | S Nat              -- Peano nats

data Fin :: Nat -> Type where     -- bounded nats
  FZ :: Fin (S n)                 --   zero
  FS :: Fin n -> Fin (S n)        --   succ

data Exp :: Nat -> Type where     -- scope-indexed
  Var :: Fin n -> Exp n           --   variables
  Lam :: Bind n -> Exp n          --   abstractions
  App :: Exp n -> Exp n -> Exp n  --   applications

data Bind n where                 -- binder type
  Bind :: Exp (S n) -> Bind n     -- increase scope
\end{lstlisting}

This snippet first defines natural numbers which can be used at the type-level,
and uses them to define bounded natural numbers (i.e., \emph{finite naturals}),
which will be used as de Bruijn indices~\cite{debruijn} to represent variables.
In terms, we use \cd{f0}, \cd{f1}, \cd{f2}, \emph{etc.} to refer to the
(bounded) numbers $0, 1, 2$, \emph{etc}, when the bound can be inferred. The
\cd{Exp} type is indexed by a natural number, the \emph{scope index}, so that it
can only represent well-scoped expressions. All variables must be in the range
specified by the scope of the datatype.  The type of the \cd{Lam} constructor
states that the body of the expression is a binding, i.e, it increases the scope
by one. For reasons which will be explained later, we create a new type
\cd{Bind} for this increase. Note that, since \cd{Fin 0} has no inhabitant,
\cd{Exp 0} cannot contain any free variable and hence represents \emph{closed
expressions}.

For example, to represent the closed lambda calculus term $\lambda x. \lambda y.
y$, we use index \cd{f0} for the occurrence of $y$ as it refers to the closest
enclosing binder.  On the other hand, for $\lambda x. \lambda y. x$, the index
of $x$ in the same scope is \cd{f1}.
\begin{lstlisting}
example1 :: Exp 0   -- \x -> \y -> y
example1 = Lam (Bind (Lam (Bind (Var f0))))

example2 :: Exp 0   -- \x -> \y -> x
example2 = Lam (Bind (Lam (Bind (Var f1))))
\end{lstlisting}

\subsection{An Environment-Based Evaluator}%
\label{sec:env-eval}

Consider the implementation of an environment-based evaluator for the
well-scoped representation. An \emph{environment}, or closing substitution, is a
mapping from variable indices to values. In the pure lambda calculus, a value is
a closure~\cite{Landin64}---an environment paired with the body of an
abstraction.

\begin{lstlisting}
type Env n = Fin n -> Val -- environment type

data Val where
  VLam :: Env n -> Bind n -> Val   -- closure
\end{lstlisting}
Environments can be constructed much like length-indexed lists.
\begin{lstlisting}
nil  :: Env Z                      -- empty env
(.:) :: Val -> Env n -> Env (S n)  -- extend w/ value
\end{lstlisting}

An environment-based evaluator uses an environment argument to remember the
values of variables.

\begin{lstlisting}
eval :: Env n -> Exp n -> Val
eval r (Var x)     = r x
eval r (Lam b)     = VLam r b
eval r (App a1 a2) = case eval r a1 of
    VLam s (Bind b) -> eval (eval r a2 .: s) b
\end{lstlisting}

The interpretation of a lambda expression is a closure. This value stores the
current environment along with the body of the lambda expression.  In the
application case, this body is evaluated with the saved environment after it has
been extended with the value of the argument of the application. Note that
Haskell's non-strict semantics gives this interpreter a call-by-need evaluation
behavior---the argument is only evaluated if the variable is used in body of the
abstraction.

There are two observations to make about this implementation.  First,
scope-safety means that the evaluator will never trigger a run-time error from
an unbound variable.  The environment type \cd{Env} uses the scope index to
statically track its domain, ensuring that every variable lookup is in scope.
Second, there is no administrative work during evaluation. Even though we are
using indices to represent variables, there is no shifting or substitution
required. Instead, everything is handled via the environment.

The fundamental mechanism of this code is the \emph{closure}, i.e., an
expression that is paired with its environment. This environment acts as a
\emph{delayed} substitution, leading to significant benefits in our
implementation.  However, despite these benefits there are also
drawbacks with this evaluator.

\begin{enumerate}[leftmargin=1em,listparindent=0pt]

\item Explicitly passing around an environment and storing a delayed
  environment in a closure doesn't look like the lambda-calculus! What if we
  wanted something that looks more like a research paper, which often use
  substitution?

\item Closures are closing substitutions, and our evaluator only works with
  closed terms. What if we wanted full normalization (i.e., reduction of open
  terms under binders) instead?

\item The result of evaluation is a closure. If we want to access the lambda
  calculus term corresponding to that closure we need to do more work, i.e.,
  apply the delayed substitution. Furthermore, when comparing the results of
  evaluation, we should not distinguish closures that differ in their saved
  environments.

\end{enumerate}

However, these issues can be readily resolved. In the next subsection, we will
take the idea of working with delayed substitutions to bring some of the
benefits of an evaluation-based interpreter to a substitution-based
implementation.

\subsection{A Substitution-Based Interpreter}%
\label{sec:subst-eval}

\begin{figure}[t]
\begin{lstlisting}
-- environment (parallel substitution) type
type Env m n = Fin m -> Exp n
-- empty and "cons"
nil    :: Env Z n
(.:)   :: Exp n -> Env m n -> Env (S m) n
-- identity and composition of environments
id     :: Env n n
(.>>)  :: Env m n -> Env n p -> Env m p
-- env that increments all variables by one
shift  :: Env n (S n)
-- lift an env to a larger scope
up     :: Env m n -> Env (S m) (S n)
up e   = Var f0 .: (e .>> shift)
\end{lstlisting}
\caption{Parallel substitutions and operations}%
\label{fig:env}
\end{figure}

\begin{figure*}[th]
\begin{tabular}{ll}
\begin{subfigure}{.45\textwidth}
\begin{lstlisting}
-- binder type
data Bind n where
  Bind :: Exp (S n) -> Bind n

-- apply a parallel substitution to a binder
applyBind :: Env m n -> Bind m -> Bind n
applyBind r (Bind b) = Bind (applyE (up r) b)

-- apply a parallel substitution to a term
applyE :: Env m n -> Exp m -> Exp n
applyE r (Var x)   = r x
applyE r (Lam b)   = Lam (applyBind r b)
applyE r (App f a) = App (applyE r f) (applyE r a)

-- single substitution
instantiate :: Bind n -> Exp n -> Exp n
instantiate (Bind b) a = applyE (a .: Var) b
\end{lstlisting}
\caption{Eager substitution}%
\label{fig:subst-eval:eager}
\end{subfigure}%
&
\begin{subfigure}{0.45\textwidth}
\begin{lstlisting}
-- binder type with delayed substitution
data Bind n where
  Bind :: Env m n -> Exp (S m) -> Bind n

-- apply a parallel substitution to a binder
applyBind :: Env m n -> Bind m -> Bind n
applyBind r (Bind r' b) = Bind (r' .>> r) b

-- apply a parallel substitution to a term
applyE :: Env m n -> Exp m -> Exp n
applyE r (Var x)   = r x
applyE r (Lam b)   = Lam (applyBind r b)
applyE r (App f a) = App (applyE r f) (applyE r a)

-- single substitution
instantiate :: Bind n -> Exp n -> Exp n
instantiate (Bind r b) a = applyE (a .: r) b
\end{lstlisting}
\caption{Delayed substitution}%
\label{fig:subst-eval:lazy}
\end{subfigure}
\end{tabular}
\caption{Eager and delayed substitutions}%
\label{fig:subst-eval}
\end{figure*}

Now consider a standard substitution-based implementation of an
interpreter for the pure lambda calculus.
\begin{lstlisting}
eval :: Exp n -> Exp n
eval (Var x)   = Var x
eval (Lam b)   = Lam b
eval (App f a) = case eval f of
     Lam b -> eval (instantiate b (eval a))
\end{lstlisting}
In this case, we don't use an auxiliary type of values. Instead, evaluation, if
it terminates, produces a new expression.  The important step is in the
application case: after evaluating the function, we substitute the evaluated
argument into the body of the lambda term before its evaluation, using the
function
\begin{lstlisting}
instantiate :: Bind n -> Exp n -> Exp n
\end{lstlisting}

The instantiation function is defined through \emph{substitution}, and a common
implementation is shown in Figure~\ref{fig:subst-eval:eager}. The definitions in
this figure rely on a small library (\pref{fig:env}) for working with mappings
from indices to expressions, also known as \emph{parallel substitutions}.

The use of \emph{parallel} substitutions means that when applying a
substitution to a lambda expression, it simultaneously replaces all free
variables in its range with expressions in the new scope. To do so, we define
a type for parallel substitutions, which we dub \emph{environments} because
they map indices in a bounded range, along with a number of operations that
construct them.  In contrast to the previous \cd{Env} type, here the type
cares about two numbers: \cd{m}, the scope of the environment (i.e., size of
its domain) and \cd{n}, the scope of the expressions in the range. This is
purely a type change; the empty and extension definitions are the same as the
previous version, but they have a new type.

The \cd{applyE} function applies the substitution to an expression. In the case
of a lambda expression, the substitution must be \emph{lifted} to work in the
increased scope, via \cd{up}. This operation modifies the substitution so that
it leaves the bound variable alone (index \cd{f0} is mapped to \cd{Var f0}), and
offsets the rest of the substitution by one, and shifts any free variables in
the range of the substitution to the new scope.

In contrast to the environment-based interpreter, working with substitutions
requires bookkeeping. This bookkeeping costs in terms of performance (both
substitution and shifting traverse the terms) and in terms of development time
(the code in Figure~\ref{fig:subst-eval:eager} is slightly longer than the one
in Section~\ref{sec:env-eval}).

However, as above, this definition is scope-safe. The type of the substitution
function tells us that it reduces the number of free variables in the term. The
type of the evaluator restricts it to working with closed
expressions.\footnote{However, unlike before, the evaluator cannot statically
guarantee that the result of evaluation (if any) will be a lambda expression, so
there is the possibility of pattern match failure.}

\subsection{A Delayed Substitution-Based Interpreter}%
\label{sec:delayed-subst}

Now let's improve our substitution-based interpreter by using ideas from the
environment-based approach. The key technique is that of a \emph{delayed
substitution}, analogous to the closure above. Instead of eagerly substituting
through the term, the term itself may contain unapplied substitutions.

One option would be to add an explicit substitution form to the expression
datatype, following the $\lambda\sigma$-calculus of Abadi et
al.~\cite{abadi:explicit-substitutions}. However, we can be a bit more sneaky,
as the only part of the term where this only really matters is at binders.

We modify the abstract type \cd{Bind}, as shown in
Figure~\ref{fig:subst-eval:lazy}, so that it also contains a delayed
environment. Note that, because we already specified the scope increase with the type \cd{Bind},
the definition of \cd{Exp} is not changed in any way. Thus, we are
smuggling an environment into our expression type, hiding it behind an abstract
type so that it does not need to be manipulated explicitly. This version of the
\cd{Bind} type generalizes both lambda expressions and closures. If the delayed
environment is \cd{id}, which maps indices to corresponding variables, then this
type is like a normal lambda abstraction. On the other hand, the type \cd{Bind
0} is like the \cd{Val} type from above and forces the delayed environment to be
a closing substitution.

With these modifications, the implementation of the evaluator is identical to
the version shown in Section~\ref{sec:subst-eval}.
\begin{lstlisting}
eval :: Exp n -> Exp n        -- same as in 2.2
eval (Var x)   = Var x
eval (Lam b)   = Lam b
eval (App f a) = case eval f of
     Lam b -> eval (instantiate b (eval a))
\end{lstlisting}

The place where the delayed substitution comes into play is in the
\cd{applyBind} operation (Figure~\ref{fig:subst-eval:lazy}). There, instead of
shifting and applying the substitution to the body of the binder, we can wait by
\emph{composing} it with the suspended substitution in the binder, using the
\cd{(.>>)} operator. This observation was already present, in a
slightly different form, in Bird and Paterson's functional
pearl~\cite{bird:debruijn}.

There is of course, no free lunch. By introducing the \cd{Bind} type, we no
longer have a unique representation for $\alpha$-equivalent lambda expressions
as they may differ in the substitutions suspended at binders. We account for
this by equating the bodies of binders only after forcing their delayed
substitutions.

\subsection{An Explicit Environment-Based Interpreter}%
\label{sec:explicit-env}

We can delay the substitution even further by explicitly passing it as an
argument to the evaluator, similar to the environment passing evaluator.  This
implementation fuses the traversal of the term during instantiation with the
traversal of the evaluator itself.

\begin{lstlisting}
eval :: Env m n -> Exp m -> Exp n
eval r (Var x)   = r x
eval r (Lam b)   = Lam (applyBind r b)
eval r (App f a) = case eval r f of
     Lam (Bind r' b) -> eval (eval r a .: r') b
\end{lstlisting}

Compared to the previous definition, this version delays substitution even
more, and ultimately does less work. With the previous version, in an
application, we evaluate the body of the binder after substituting its
(evaluated) argument for its parameter. That means that the (evaluated)
argument gets re-evaluated again for each occurrence of the variable in the
original body. Re-evaluation is fast, as this argument is already a value, but
the revised version avoids it entirely.

\subsection{What Is the Point of All of This?}%
\label{sec:eval-benchmarks}

In this section, we have considered four different evaluators (called
\texttt{EvalV},\texttt{SubstV}, \texttt{BindV} and \texttt{EnvV} respectively).
These implementations are straightforward and directly map to our understanding
of syntactic manipulations of the lambda calculus.  However, these four
evaluators perform differently, as shown by the table below. In each case, we
timed the evaluation of the same large expression.\footnote{The large expression
was developed by Lennart Augustsson~\cite{augustsson:cooked} and is the Scott
encoding of \cdfoot{fact 6 == sum [0.. 37] + 17}.\ifauthor{} The term is shown
in Appendix~\ref{app:lennartb}.\fi{} To observe the result of evaluation, we
included the boolean values \cdfoot{true} and \cdfoot{false} in the language and
extended the evaluators accordingly. The benchmarks were run on a 2024 MacBook
Pro M4 with 48 GB memory. Reported times are OLS estimates computed using the
\cdfoot{criterion} library. These benchmarks are available in the directory
\cdfoot{benchmark/lib/Rebound/Manual/Lazy}.}

\begin{table}[H] 
\begin{tabular}{llll}
  \toprule
  Name & Detailed in & Description & Time\\
  \midrule
  EvalV & \pref{sec:env-eval} & Original & 0.286 ms\\
  SubstV & \pref{sec:subst-eval} & Standard subst. & 3330 ms\\
  BindV & \pref{sec:delayed-subst} & Delayed subst. & 0.767 ms\\
  EnvV & \pref{sec:explicit-env} & Env. argument & 0.586 ms\\
  ExpSubstV & (not shown) & Explicit subst & 5.96 ms\\
  \bottomrule
\end{tabular}
\end{table}

The first line of the table (\cd{EvalV}), the pure environment-based evaluator,
is our baseline and produces the fastest time. The version with direct
implementation of substitution (\cd{SubstV}) is orders of magnitude slower.
Notably, delaying substitutions in \cd{Bind} (\cd{BindV}) is enough to recoup
most of the lost time: it takes us back to a time in the same order of magnitude
as \cd{EvalV}. Passing the environment explicitly (\cd{EnvV}) brings us to about
twice as slow as the original version. (But note that these two versions also
work with open terms, unlike the original evaluator.) We also compared these
implementations with a fifth version, \texttt{ExpSubstV}, that more closely
resembles the $\lambda\sigma$-calculus~\cite{abadi:explicit-substitutions} and
allows suspended substitutions anywhere in the abstract syntax. However, on this
benchmark that implementation is about ten times slower than \cd{EnvV}.

\section{The \rebound{} Library}%
\label{sec:library}

We don't need to start from scratch in our next implementation of a language
with binding. In this section, we separate the mechanism from the previous
section into parts that are specific to the untyped lambda calculus and parts
that can be reused for other languages and purposes and package that up in the
\rebound{} library.

\begin{figure}[b]
\begin{lstlisting}
-- delayed substitution (abstract type)
type Env v m n
-- access environment at index m
(!) :: Env v m n -> Fin m -> v n
-- operations from Figure 1:
nil, (.:), id, (.>>), shift, up
-- identify the variable constructor
class Subst v v => SubstVar v where
  var  :: Fin n -> v n
-- apply environment to a term
class SubstVar v => Subst v e where
  applyE :: Env v m n -> e m -> e n
-- bind var v in body e (abstract type)
data Bind v e n
-- `Subst' instance for `Bind` (i.e., applyBind)
instance SubstVar v => Subst v (Bind v e)
-- single substitution
instantiate :: Bind v e n -> v n -> e n
\end{lstlisting}
\caption{Core library interface}%
\label{fig:library}
\end{figure}

\begin{figure}
\begin{lstlisting}
-- scope-indexed syntax
data Exp :: Nat -> Type where
  Var :: Fin n -> Exp n
  Lam :: Bind Exp Exp n -> Exp n
  App :: Exp n -> Exp n -> Exp n
    -- (==) tests alpha-equivalence
    deriving (Eq)

-- identify the variable constructor
instance SubstVar Exp where var = Var

-- direct implementation of substitution
instance Subst Exp Exp where
  applyE r (Var x)   = r ! x
  applyE r (Lam b)   = Lam (applyE r b)
  applyE r (App f a) = App (applyE r f) (applyE r a)

-- implementation with generic programming
instance Subst Exp Exp where
  isVar (Var x) = Just (Refl, x)
  isVar _ = Nothing
\end{lstlisting}
\caption{User code for well-scoped terms}%
\label{fig:exp-instance}
\end{figure}

Figure~\ref{fig:library} isolates the library definitions necessary to implement
the evaluation functions from Section~\ref{sec:examples}.  Then,
Figure~\ref{fig:exp-instance} uses these operations to implement substitution
for the untyped lambda calculus twice: first directly and then by deriving the
definition using generic programming.

This first \cd{Subst} instance is simple because the library already includes an
instance for applying the substitution to a binder: the composition and delaying
of the \cd{Env} type happens behind the scenes.
The second instance only requires the user to identify the variable case
in the abstract syntax (if there is one), but requires no modification when new
syntactic forms are added to the language.

\rebound{} keeps the \cd{Env} type abstract. While one way to implement delayed
substitutions is with functions, as shown in the previous section, that is not
the only possible implementation. We discuss alternatives in
Section~\ref{sec:optimizations}. Because this type is abstract, we include an
explicit operator \cd{!} for looking up an index in the environment.

The \cd{SubstVar} class identifies scope-indexed types that have variable
constructors.  The \cd{Subst} type class takes two arguments. The first, \cd{v},
describes the co-domain of the deferred substitution (i.e., what type do
variables stand for) and the second \cd{e} describes the type we are
substituting into.  Often, these two types will be the same, e.g., in
Figure~\ref{fig:exp-instance}, we instantiate both parameters of the \cd{Subst}
class with \cd{Exp}.

\paragraph{The abstract \cd{Bind} type}

The type of single binders (\cd{Bind}) is abstract and the library includes
relevant type class instances for this type, such as \cd{Subst}. Internally, the
\cd{Bind} type includes a suspended environment, as in
Figure~\ref{fig:subst-eval:lazy}, but users need not be aware of this delayed
substitution.  Instead, they should work with the \cd{bind} and \cd{getBody}
wrappers.

\begin{lstlisting}
bind :: SubstVar v => e (S n) -> Bind v e n
bind = Bind id

getBody :: Subst v e => Bind v e n -> e (S n)
getBody (Bind r e) = applyE (up r) e
\end{lstlisting}

\rebound{} also includes operations that allow users to manipulate environments
explicitly. For example, a user may wish to instantiate a binder while calling a
function that is parameterized by the current environment.

\begin{lstlisting}
instantiateWith :: Bind v e n -> v n
                   -> (forall m. Env v m n -> e m -> d n)
                   -> d n
instantiateWith f (Bind r a) v = f (v .: r) a
\end{lstlisting}

This library function is exactly what is required to implement the
environment-based interpreter shown in Section~\ref{sec:explicit-env} while
keeping the \cd{Bind} and \cd{Env} types abstract.

\subsection{Beyond the Untyped Lambda Calculus}

Many languages include rich binding structures. We would also like to implement
more functions than evaluators, such as normalizers (which reduce open terms)
and type checkers.  Finally, we would like to use this library in full-featured
implementations, so it must be compatible with their additional requirements.

To demonstrate the features of this library, we have used it to represent the
binding structure for a number of different calculi, and have implemented
normalizers and type checkers for these languages. These examples have been
extensively documented and are distributed along with the library.

\begin{description}[leftmargin=0pt,listparindent=0pt]
\item[\texttt{LC.hs}] Untyped lambda calculus with single binding. Big-step and
small-step evaluation functions using substitution, normalization.
\item[\texttt{LClet.hs}] Untyped lambda calculus with let binding, which may be
recursive or nested. Big-step evaluation and normalization.
\item[\texttt{Pat.hs}] Untyped lambda calculus with constants and pattern
matching. Big-step and small-step evaluation.
\item[\texttt{SystemF.hs}] System F with separate term and type variables. Type
checker.
\item[\texttt{PureSystemF.hs}] System F with a unique syntactic class for terms
and types. Type checker.
\item[\texttt{PTS.hs}] Dependently-typed calculus including $\Pi$ and weak
$\Sigma$ types, based on Pure Type Systems~\cite{barendregt:pts}. Big-step and
small-step evaluation, normalization. Bidirectional type checker.
\item[\texttt{DepMatch.hs}] Dependently-typed calculus with nested dependent
pattern matching for strong $\Sigma$ types. Big-step and small-step evaluation,
normalization. Bidirectional type checker.
\end{description}
We also have a few examples that demonstrate how to work with well-scoped
expressions.
\begin{description}[leftmargin=0pt,listparindent=0pt]
\item[\texttt{ScopeCheck.hs}] Scope checker: converts a nominal representation
  of binding to a well-scoped version.
\item[\texttt{LCQC.hs}] Generator for well-scoped untyped lambda calculus terms,
  suitable for property-based testing using the
  QuickCheck~\cite{claessen:quickcheck} library.
\item[\texttt{HOAS.hs}] Uses HOAS as a convenient interface to construct
  concrete well-scoped expressions.\footnote{This example is inspired by
  McBride's classy hack:
  \url{https://mazzo.li/epilogue/index.html\%3Fp=773.html} }
\item[\texttt{PatGen.hs}]
  A version of \cd{Pat.hs} that demonstrates the use of generic programming
  in the presence of sophisticated scoped-indexed types.
\end{description}

\subsection{Pattern Binding}%
\label{sec:patterns}

Single binders work well for theoretical developments. But we often want more
from a binding library in a practical implementation, such as \emph{pattern
binding}. A pattern can be any type: all we need to know about it is how many
variables it binds.  Figure~\ref{fig:pat} shows the generalized interface for
the \cd{Bind} type from a single index to \emph{pattern binding}.
\begin{figure}
\begin{lstlisting}
class Sized (t :: Type) where
  -- retrieve size from the type (number of variables
  -- bound by the pattern)
  type Size t :: Nat
  -- access size as a natural number term
  size :: t -> SNat (Size t)

-- bind variables for v, in expressions c
-- with patterns p in scope n
type Bind v c (p :: Type) (n :: Nat)

-- create a binder for the pattern p, introducing
-- its variables into the scope
bind  :: (Sized p, Subst v c) => p -> c (Size p + n)
      -> Bind v c p n

-- instantiate a binder by filling in values for
-- the variables bound by the pattern
instantiate  :: (Sized p, Subst v c) => Bind v c p n
             -> Env v (Size pat) n -> c n
\end{lstlisting}
\caption{Pattern binding interface}%
\label{fig:pat}
\end{figure}

The type class \cd{Sized} describes types that statically identify the number of
variables that they bind, using the associated type \cd{Size}.  This class also
includes the function \cd{size} that returns the same information as a
\emph{singleton type}~\cite{eisenberg:singletons}. The type \cd{SNat n} contains
a natural number isomorphic to \cd{n}. Because we lack true dependent types in
Haskell, singleton types provide a bridge between runtime and compile-time data.

\paragraph{$n$-ary binding}
The simplest form of pattern binding, is binding several variables at once. For
example, the language of \texttt{PTS.hs} eliminates products using pattern
matching instead of projections. It includes a ``split'' term that
simultaneously binds \emph{two} variables to the two components of the pattern.
Therefore, it uses the singleton type \cd{SNat 2} as a pattern that binds
exactly two variables.

\begin{lstlisting}
data Exp n where
  ... -- other constructors as before
  -- create a product `(e1, e2)`
  Pair  :: Exp n -> Exp n -> Exp n
  -- split a product `let (x,y) = e1 in e2`
  -- the body of the binder has extended scope (2 + n)
  Split :: Exp n -> Bind Exp Exp (SNat 2) n -> Exp n
\end{lstlisting}

Because the number of bound variables can be statically determined from the
pattern, the \cd{Bind} constructor in Figure~\ref{fig:pat} increases the scope
of the body of the binder by the number of variables bound in the pattern, and
requires the same number of values in instantiation. Continuing this example, we
extend the evaluator with cases for \cd{Pair} and \cd{Split} as below. In the
latter case, the type checker requires us to supply two arguments to
\cd{instantiate}, packaged in an environment.

\begin{lstlisting}
eval (Pair a1 a2)  = Pair a1 a2
eval (Split a b)   = case eval a of
    Pair a1 a2 ->
      eval (instantiate b (a1' .: a2' .: nil))
        where a1' = eval a1
              a2' = eval a2
\end{lstlisting}

\paragraph{Nested pattern matching}

Pattern binding also extends to arbitrary datatype patterns and nested pattern
matching. For example, suppose we would like to add the ability to deeply match
tuples in let bindings, i.e., \cd{let (x, (y, z)) = e1 in e2}. To do so, we can
define a datatype to represent the tuple structure of the pattern (\cd{Pat}) and
use this pattern in a new expression form (\cd{LetPair}).

\begin{lstlisting}
data Pat (m :: Nat) where
  PVar  :: Pat N1 -- binds a single variable
  PPair :: Pat m1 -> Pat m2 -> Pat (m2 + m1)

data Exp (n :: Nat) where
  ... -- other constructors as before
  LetPair :: Exp n -> Bind Exp Exp (Pat m) n  -> Exp n
\end{lstlisting}

Above, a pair pattern is either a single variable or an application of the
pair constructor to two nested patterns. To know how many arguments are bound
in this case, we sum the number of binding variables in each subpattern.

The evaluator for \cd{LetPair} expressions must first identify whether the
pattern matches a given value, and if so, produce a substitution for each of the
variables in the pattern to the corresponding subterms in the
value.\footnote{For simplicity, the code above throws an error when the
expression has the wrong form; a more realistic example would gracefully handle
this case.}
\begin{lstlisting}
eval (LetPair a b) = eval (instantiate (getBody b) r)
  where
    r = patternMatch (getPat b) (eval a)

patternMatch :: Pat p -> Exp m -> Env Exp p m
patternMatch PVar e = e .: nil
patternMatch (PPair p1 p2) (Pair e1 e2) =
  withSNat (size p2) (r2 .++ r1) where
    r1 = patternMatch p1 e1
    r2 = patternMatch p2 e2
\end{lstlisting}
For \cd{PPair}, we use an environment append operation, \cd{(.++)}, to combine
the results of pattern matching the components of the pair. This operation
implicitly needs the length of its first argument at runtime; the function
\cd{withSNat} uses a value of type \cd{SNat n} to satisfy this constraint.

This idea can also be used to implement pattern matching for arbitrary
datatypes, as we demonstrate in the example \cd{Pat.hs}.

\section{Case Study: \piforall{}}%
\label{sec:pi-forall}

To test the expressiveness of our library, we have ported
\piforall~\cite{weirich:pi-forall}, a demo implementation of a type checker for
a dependently typed programming language, to \rebound{}. The \piforall{}
implementation includes a parser and type checker for a language with dependent
functions, datatypes, dependent pattern matching, multiple modules and
informative error messages.  The original implementation used the
\texttt{unbound-generics}~\cite{unbound-generics,weirich:binders} library
(called \unbound{} for short) to implement substitution and alpha-equivalence.
This binding library relies on a \emph{locally nameless representation}.

The previous implementation of \piforall{} did not statically track the scoping
of variables, relying instead on \unbound{}'s design to ensure a correct
treatment of binders.  Therefore, we were curious to learn whether \rebound{},
and more generally intrinsically scoped representations, could be used in a
setting that is closer to a practical implementation. While \piforall{} is a
tutorial, focusing more on explaining how dependent types work than on
developing a robust and efficient language, the features of \piforall{} make it
more than just a toy example.

The goal of this re-implementation is to evaluate the expressiveness of the
core library by using it to implement a non-trivial programming language. As
part of this process, it provided practical motivation for new extensions of
the library.  In particular, two features of \piforall{} provide the greatest
challenge to
\rebound{}; we discuss them below.

\paragraph{Error messages:}
Feedback from the type-checker to the user is crucial in dependently-typed
languages as the types become expressive/complex.  This feedback comes in the
form of type errors, warnings, and a special \texttt{PRINTME} expression that
instructs the type-checker to print the types of all variables currently in
scope.
It is important for such feedback to refer to variables as they were defined
by the user and not using their index in the (current) scope. This means that
\piforall{}'s type-checker must maintain a mapping from de Bruijn indices to
user-defined names during all parts of the implementation.

\paragraph{Datatypes and pattern matching:}

A significant amount of code in the \piforall{} type checker involves checking
datatype declarations, uses of type constructors and data constructors, and
pattern matching expressions.  Supporting \piforall{}'s indexed datatypes
requires expressive support for \emph{telescopes}, sequences of variable
declarations where the type of each identifier may refer to any variable bound
earlier in the telescope. Telescopes are complex patterns, as they bind
variables both internally (later in the same telescope) and externally (in the
subsequent expression).  Scope safety means that two values must be tracked: the
number of variables bound by the telescope, and the (extending) scope for its
embedded expressions.

\subsection{Scoped Monads}

When working with abstract syntax, Haskell programmers often use a
\texttt{Reader} monad~\cite{jones:1995} to store information about in-scope
variables, such as user-supplied names, types, or definitions. However, when
working with de Bruijn indexed terms and statically tracking the current scope,
the usual reader monad is not sufficiently expressive. For example, when storing
a length-indexed vector of the names of variables currently in scope, we might
like to define an operation for extending that scope (i.e., consing a new name
to the vector).
\begin{lstlisting}
-- a simple monad that tracks names currently in scope.
-- Note that the scope is part of the type!
type Scoped n = Reader (Vec String n)
-- a specialized version of local
addToContext :: String -> Scoped (S n) a -> Scoped n a
addToContext x = local /&(x :>)&/ -- type error!
\end{lstlisting}
The expression \cd{(x :>)} has type
\begin{lstlisting}
Vec String n -> Vec String /&(S n)&/
\end{lstlisting}
but \cd{local}'s type requires a function with type
\begin{lstlisting}
Vec String n -> Vec String /&n&/
\end{lstlisting}

Therefore, to maintain this information, \rebound{} defines the
\cd{ScopedReaderMonad} class. As its name implies, this monad offers the same
API which allows to read and update a piece of data, usually called the monad's
\emph{environment}. The key difference is in the types: the
\cd{ScopedReaderMonad}'s environment has to be indexed by a scope, and the
\cd{localS} operation is allowed to change the scope.
\begin{lstlisting}
class (forall n. Monad (m n))
  => MonadScopedReader (e:: Nat -> Type) m | m -> e
where
  -- retrieves the monad environment
  askS :: m n (e n)
  -- executes a computation in a modified environment
  localS :: (e n -> e n') -> m n' a -> m n a
  -- retrieves a function of the current environment
  readerS :: (e n -> a) -> m n a
\end{lstlisting}

By defining the \cd{Scoped} monad so that it is an instance of this class,
the \cd{localS} method has the type that we need.
\begin{lstlisting}
instance MonadScopedReader (Vec String) Scoped
  where ...

addToContext :: String -> Scoped (S n) a -> Scoped n a
addToContext x = localS /&(x :>)&/ -- type checks!
\end{lstlisting}

\subsection{Scoped Patterns and Telescopes}

The most significant issue with datatype definitions in \piforall{} is that the
telescopes for constructors both bind new variables and include occurrences of
existing variables.  For example, the usual length-indexed vector can be
expressed in \piforall{} using the following top-level declaration.
\begin{lstlisting}
data Vec (A : Type) (n : Nat) : Type where
  Nil  of [n = Zero]
  Cons of [m : Nat] (h : A) (t : Vec A m) [n = Succ m]
\end{lstlisting}

This declaration includes a telescope for the parameters of the \cd{Vec} type
(i.e., \cd{A} and \cd{n}) and a telescope (in the scope of the first one!) for
the parameters of each constructor (e.g., \cd{m}, \cd{h}, and \cd{t} for
\cd{Cons}). The telescopes for constructors may also include constraints (or
``Ford equations''~\cite{mcbride:thesis}) on the parameters, such as \cd{n =
Zero} in the \cd{Nil} case, constraining the length to be \cd{Zero} for empty
vectors.

This dual treatment of variables means that the simple datatypes for patterns,
presented in \pref{sec:patterns}, are not expressive enough. Instead, we need to
statically track both the number of bound variables and the current scope. In
other words, we use patterns of kind \cd{Nat -> Nat -> Type} instead of \cd{Nat
-> Type}, where the first argument is the number of bound variables and the
second argument is the current scope regulating free variables.

Furthermore, in a binding telescope, variables bound earlier in the telescope
can occur in types and constraints that appear later in the telescope. Re-using
\cd{Cons} as an example, its telescope binds the variable \cd{m} and then uses
it as both the length of the sublist \cd{t}, and in the constraint on \cd{n}.

The \rebound{} library defines the \cd{TeleList} datatype to support telescopes.
A \cd{TeleList} is parameterized by both \cd{p}, the number of variables that it
binds and \cd{n} the scope that it appears in. It is also generic over \cd{pat},
a similarly parameterized pattern type for each entry in the telescope. In the
\cd{TNil} case, the telescope binds no variables \cd{N0} and is available in any
scope (\cd{n} is unconstrained). However, in the \cd{TCons} case, if the entry
binds \cd{p1} variables, then the rest of the telescope occurs in the extended
scope \cd{p1 + n}. Furthermore, the number of variables bound by the telescope
includes both those bound here in the head and those bound later in the tail.
\begin{lstlisting}
data TeleList (pat :: Nat -> Nat -> Type) p n where
  TNil  :: ( ... ) => TeleList pat N0 n
  TCons :: ( ... ) =>
    pat p1 n -> TeleList pat p2 (p1 + n) ->
    TeleList pat (p2 + p1) n
\end{lstlisting}

Using this generic definition, \piforall{}'s telescopes can be defined by first
a \cd{Local} type describing an element of the telescope, and then applying
\cd{TeleList} to it:
\begin{lstlisting}
data Local p n where
  -- Variable binding, e.g., (h: A)
  LocalDecl :: LocalName -> Typ n -> Local N1 n
  -- "Ford" constraint, e.g., [n = Succ m]
  LocalDef  :: Fin n -> Term n -> Local N0 n

type Telescope = TeleList Local
\end{lstlisting}
This type represents either a local variable declaration or an equality
constraint. In the former, the pattern binds one variable, and the type of that
variable is in scope \cd{n}. In the latter, no variables are bound, but the
equation must have a variable in the current scope on the left-hand side, and a
term in the current scope on the right hand side.

The constructors also includes constraints (elided) that help Haskell's type
checker work with telescopes. For \cd{TNil}, the constraint states that $n + 0 =
n$. The \cd{TCons} constructor has two. The first states that the size of the
pattern is independent of the scope in which it appears. More formally,
\begin{center}
  \cd{forall n. Size (pat p n) ~ p}
\end{center}
The second asserts an associativity property about addition, instantiated with
the binding variables. The \rebound{} library includes smart constructors to
supply these two constraints automatically. These facts are brought into scope
whenever the telescope is pattern matched, so are automatically available to the
type checker during traversal of the telescope.

\section{Benchmarks}%
\label{sec:benchmarks}

Here we justify our claim that \rebound{} provides an efficient implementation
of interpreters and type checkers. As we report in this section, we have
developed two sorts of benchmarks: \emph{normalization} and \emph{dependent type
checking}.

The normalization benchmarks are a broad comparison across multiple
implementations of lambda calculus normalization. We use these benchmarks to
compare different ways of using \rebound{}, different libraries for binding
(\pref{sec:nf-benches}), and different implementations of \rebound{}'s
environment data structure (\pref{sec:optimizations}).
The dependent type checking benchmarks (\pref{sec:tc-benches}) compare the
performance between two versions of \piforall{}. They model a more realistic
language and draw on several different operations on syntax working together in
a more realistic usage.

\subsection{Normalization Benchmarks}%
\label{sec:nf-benches}

\begin{table}
\caption{Comparison of normalization benchmarks}
\begin{tabular}{llll}
  \toprule
  Benchmark Name & eval & nf & random15\\
  \midrule
  Env.Strict.BindV & 1.01 ms & 1.21 ms & 0.624 ms\\
  Env.Strict.EnvV & 0.645 ms & 0.868 ms & 0.523 ms\\
  Env.Strict.EnvGenV & 0.777 ms & 1.24 ms & 0.728 ms\\
  Env.Strict.Bind & 4.26 ms & 4.39 ms & 0.593 ms\\
  Env.Strict.Env & 0.674 ms & 0.91 ms & 0.531 ms\\
  Env.Strict.EnvGen & 0.804 ms & 1.28 ms & 0.77 ms\\
  \midrule
  DeBruijn.BoundV & 1.07 ms & 1.19 ms & 3.77 ms\\
  DeBruijn.Bound & 4.03 ms & 4.14 ms & 3.67 ms\\
  Named.Foil & 167 ms & 169 ms & 194 ms\\
  Unbound.Gen & 1830 ms & 1770 ms & 16.7 ms\\
  Unbound.NonGen & 1160 ms & 1110 ms & 3.02 ms\\
  \midrule
  NBE.KovacsScoped & 0.329 ms & 0.333 ms & 0.0846 ms\\
  \bottomrule
\end{tabular}%
\label{table:comparison}
\end{table}

This section compares implementations of normalization for the untyped lambda
calculus expressions. Our normalization function, \cd{nf} fully reduces its
argument, including underneath binders. It is defined in terms of an auxiliary
function \cd{whnf}, that calculates the \emph{weak-head normal form} of an
expression, i.e., reduces just enough to reveal the top-level structure. \ifauthor
The implementation of these functions appears in 
Appendix~\ref{app:nf}. \fi

Table~\ref{table:comparison} shows the results
on various tasks.
\begin{description}[leftmargin=0pt,listparindent=0pt]
\item[\texttt{eval}] Weak-head reduction of Augustsson's term encoded in the
untyped lambda calculus extended with boolean values. This is the same benchmark
used in Section~\ref{sec:eval-benchmarks}.
\item[\texttt{nf}] Full normalization of Augustsson's original term.
\item[\texttt{random15}] Full normalization of a collection of 100 randomly
  generated terms that need \emph{at least} \texttt{15} steps to normalize.
\end{description}

\paragraph{\rebound{} implementations}

The first six lines of Table~\ref{table:comparison} are implementations of
full reduction using \rebound. These include \texttt{Env.Strict.Bind}, the
analogue to the delayed substitution implementation of
Section~\ref{sec:delayed-subst}, and \texttt{Env.Strict.Env}, the analogue to
the explicit environment implementation of Section~\ref{sec:explicit-env}. For
this section, we do not include analogous
of \texttt{Env.Lazy.EvalV}, \texttt{Env.Lazy.SubstV}
or \texttt{Env.Lazy.ExpSubstV}. We omit the first because we would like to
measure full normalization, which is not supported by that interpreter. The
latter two are omitted because they are orders of magnitude slower during
evaluation.

In this benchmark set, it makes a (small but measurable) difference whether
the abstract syntax trees used to represent lambda calculus terms are strict
or non-strict. In Section~\ref{sec:examples}, we used a non-strict
representation for simplicity.  Here, for uniform comparison, we exclusively
use strict abstract syntax (both for \rebound{} and other implementations).

We also explore the impact of reducing the argument before beta-reduction. In
other words, whether we instantiate in both the \cd{nf} and \cd{whnf} functions
with \cd{a} or with \cd{whnf a}.  The names of benchmarks that use \cd{whnf a}
end with \cd{V}. In \cd{nf}, this modification is practical only when \cd{a} is
evaluated lazily, as normalizing all subexpressions can cause significant
blow-up.

Finally, we modified the explicit environment versions (resulting in
\texttt{Env.Strict.EnvGenV} and \texttt{Env.Strict.EnvGen}) to use
\texttt{GHC.Generics} so that we can measure the cost of generic programming.
Roughly, we found a 20--45\% cost for this convenience.

\paragraph{Other implementations}

To compare how \rebound{} stacks up, we compared its performance against several
other libraries (Section~\ref{sec:related-work} describes these libraries in
more detail).
\begin{description}[leftmargin=0pt,listparindent=0pt]
\item[\texttt{DeBruijn.Bound}] defines well-scoped de Bruijn indices using
Edward Kmett's \Bound{} library~\cite{kmett:bound}.
\item [\texttt{Named.Foil}] uses a nominal representation of lambda calculus
terms. This code was developed by the \texttt{foil} library
authors~\cite{maclaurin:foil}.
\item[\texttt{Unbound.Gen}] uses generic programming via
\unbound{}~\cite{weirich:binders,unbound-generics}. The version
\texttt{Unbound.NonGen} defines relevant operations by hand.
\end{description}

The results appear in the bottom half of Table~\ref{table:comparison}. Overall,
the \rebound{}-based implementation \texttt{Env.Strict.EnvV} is the fastest. The
\Bound{}-based implementations are competitive for \cd{eval} and \cd{nf}, but
significantly slower on the random terms. Because the \cd{Foil} and \unbound{}
versions do not delay substitutions, they are significantly slower on all
benchmarks.

The bottom of the table includes a normalization function developed by Kov\'acs
(and modified to use well-scoped expressions by the
authors).\footnote{\url{https://github.com/AndrasKovacs/elaboration-zoo/blob/master/01-eval-closures-debruijn/Main.hs}}
This version does not use the same algorithm---instead it uses
\emph{normalization-by-evaluation}, an alternative that is not based on
substitution, and so does not generalize to other operations on lambda-calculus
terms such as type checking or compiler optimization. We include it as baseline
comparison with a fast algorithm, and it is consistently the fastest version.
This suggests that when performance is critical, programmers can still use
well-scoped representations, but may wish to look for specialized algorithms for
the normalization part of their code base.

\subsection{Environment Implementation}%
\label{sec:optimizations}

\begin{table}
\caption{Comparison of different environments}
\begin{tabular}{llll}
  \toprule
  Benchmark Name & eval & nf & random15\\
  \midrule
  main/Functional & 1.11 ms & 126 ms & 0.747 ms\\
  main/Lazy & 0.632 ms & 0.84 ms & 0.492 ms\\
  main/LazyA & 0.59 ms & 0.844 ms & 0.483 ms\\
  main/LazyB & 1.52 ms & 2.36 ms & 0.795 ms\\
  main/Strict & 0.695 ms & 0.87 ms & 0.56 ms\\
  main/StrictA & 0.689 ms & 0.936 ms & 0.544 ms\\
  main/StrictB & 1.62 ms & 2.41 ms & 0.793 ms\\
  nat-word/Lazy & 0.67 ms & 0.997 ms & 0.54 ms\\
  vector/Vector & 2.69 ms & 1230 ms & 1.94 ms\\
  \bottomrule
\end{tabular}%
\label{table:ablate}
\end{table}

Figure~\ref{fig:library} defines the interface for the \rebound{} library
using an abstract type for environments (\cd{Env}). As part of our
benchmarking, we compared \texttt{Env.Strict.EnvV} compiled with several
different implementations for this data structure.
The results are shown in Table~\ref{table:ablate}.

\begin{description}[leftmargin=0pt,listparindent=0pt]
\item[\texttt{main/Functional}] is the simplest implementation and represents
  environments as functions of type \cd{Fin n -> Exp m}.
\item[\texttt{main/Lazy}] is the implementation that we use in the library (and
  for the benchmarks in Table~\ref{table:comparison}). This implementation
  ``defunctionalizes'' the environment as a data structure, representing
  environment creation functions (\cd{idE}, \cd{(.:)}, \cd{shift}, etc) as
  constructors. As a result, operations such as composition can perform
  optimizations, such as those found in Abadi et
  al.~\cite{abadi:explicit-substitutions}. Furthermore, applications of the
  identity substitution can be optimized away~\cite{wadler:explicit-weakening}.
\item[\texttt{main/LazyA}] This version is the same as \cd{main/Lazy}, but does
  not include the optimized identity application.
\item[\texttt{main/LazyB}] This version is the same as \cd{main/Lazy}, but does
  not include the optimized construction.
\item[\texttt{main/Strict},\texttt{main/StrictA},\texttt{main/StrictB}] These
  versions are the same as \cd{main/Lazy} and its variants, but use a strict
  spine.
\end{description}

These benchmarks reveal that the defunctionalized and optimized version is
faster than using a function to represent the environment, and that much of
the speed up comes from ``smart composition''. Indeed, in the lazy version,
the optimized identity application slightly degrades performance. The strict
variants are also slightly slower.

Note that none of the environments is strict in the substituted values. This is
critical as the amount of memory used to store substitutions can very easily
blow up. In order to make a fully strict substitution practical, one would have
to use other techniques to prune the stored
substitutions~\cite{abel:ordered,flint}. Of course, laziness does not completely
preclude such memory blowups, but we never encountered such an issue in our
benchmarks.

The next benchmark (\texttt{\bf nat-word/Lazy}) replaces runtime natural numbers
(\texttt{SNat} and \texttt{Fin}) with machine words in \texttt{main/Lazy}.
Finally, \texttt{\bf vector/Vector} represents environments using
\cd{Data.Sequence} in addition to using machine words.

Overall, and somewhat surprisingly, the performance for the version with machine
words is slightly worse than its unary analogue, and the results for
\texttt{Data.Sequence} are significantly worse than our original version. There
could be multiple factors at play. The \texttt{vector/Vector} implementation
cannot take advantage of the optimizations of the defunctionalized version.
Furthermore, while this data structure remains lazy in the elements stored in
the environment, the structure of the vector is computed strictly. As a result,
representing a shifting environment in a large scope takes much more space.
Finally, this version needs a runtime representation of the current scope
consistently throughout, requiring additional overhead.

Since there are tradeoffs in optimizing the identity substitution, the library
allows users to decide for themselves whether their substitution should use this
optimization or not.\@ \rebound{} provides an \cd{applyOpt} function which
performs this additional optimization. The library never uses it internally,
except for \cd{Bind}, to ensure that \cd{unbind (bind t)} returns \cd{t} in
constant time.

\subsection{Type Checking Benchmarks}%
\label{sec:tc-benches}
\begin{table*}
\caption{Comparison of the two \piforall{} implementations}
\begin{tabular}{lrrrrrrrrrr}
  \toprule
   & \multicolumn{2}{c}{AVL} & \multicolumn{2}{c}{DepAVL} & \multicolumn{2}{c}{Compiler} & \multicolumn{2}{c}{Lennart} & \multicolumn{2}{c}{CompCk}\\
  \cmidrule(lr){2-3} \cmidrule(lr){4-5} \cmidrule(lr){6-7} \cmidrule(lr){8-9} \cmidrule(lr){10-11}
   & \multicolumn{1}{c}{Time} & \multicolumn{1}{c}{Space} & \multicolumn{1}{c}{Time} & \multicolumn{1}{c}{Space} & \multicolumn{1}{c}{Time} & \multicolumn{1}{c}{Space} & \multicolumn{1}{c}{Time} & \multicolumn{1}{c}{Space} & \multicolumn{1}{c}{Time} & \multicolumn{1}{c}{Space}\\
  \midrule
  Unbound & 25.4 ms & 354 MB & 44.8 ms & 616 MB & 25.6 ms & 356 MB & 1780 ms & 20037 MB & 1610 ms & 30198 MB\\
  Rebound & 20 ms & 259 MB & 34.2 ms & 448 MB & 21.6 ms & 288 MB & 45.1 ms & 554 MB & 176 ms & 2435 MB\\
  \bottomrule
\end{tabular}%
\label{table:pi}
\end{table*}

The next set of benchmarks compares the new implementation of \piforall{},
described in \pref{sec:pi-forall}, with the original implementation.  These
benchmarks take the form of short (at most 250 lines) \piforall{} programs. The
time shown in Table~\ref{table:pi} is the end-to-end time to process the file.
We also use this set of benchmarks to take a look at \rebound{}'s memory
usage\footnote{Note that time and space usage were benchmarked separately, as
the instrumentation required to measure space has a detrimental effect on the
run time.}.
\begin{description}[leftmargin=0pt,listparindent=0pt]
\item[\texttt{AVL}] An AVL tree implementation. This implementation is standard
and doesn't rely on dependent-types.
\item[\texttt{DepAVL}] An AVL tree implementation that internally uses
dependent-types to enforce the AVL invariants.
\item[\texttt{Compiler}] A compiler from intrinsically typed arithmetic
expressions to an intrinsically typed stack language, plus interpreters for both
languages.
\item[\texttt{Lennart}] An adaptation of the benchmark from
\pref{sec:eval-benchmarks}.
\item[\texttt{CompCk}] Checks that directly interpreting a program
computing the factorial of 8 and interpreting the compiled program yield the
same result.
\end{description}
According to these benchmarks, the \rebound{}-based implementation is both
faster and allocates less memory. The gap is most stark on the ``compute
intensive'' benchmarks, but the difference is noticeable across the board.

The last two benchmarks could be considered synthetic, as they are not programs
one would typically write. However they demonstrate that heavy computations can
occur in dependently-typed languages at the type-level. Furthermore, it is
reassuring to observe that the performance difference between \rebound{} and
\unbound{} also occurs in a more fleshed out setting.

It is interesting that the space ratio between \unbound{} and \rebound{} is
always in the same order of magnitude as the time ratio. This could hint that,
in a garbage-collected and lazy language, time and space consumption tend to go
hand in hand.

\section{Is Scope-Safety a Win?}%
\label{sec:discussion}

\rebound{} uses a \emph{scoped} representation of terms, where the current scope
is tracked statically by the type system. As a result, the operations supported
by this library have expressive types, which can help users avoid bugs.

However, there is a cost associated with working with a scope-safe
representation in terms of development time and flexibility. While more expert
users, familiar with dependently-typed programming in Haskell, may benefit from
the enhanced static checking, novice Haskell programmers may struggle with the
complexity of the interface. Furthermore, scope safety may not always be the
most appropriate representation choice, due to the limitations that we list
below.

\begin{description}[leftmargin=0pt,listparindent=0pt]

\item[Reasoning about natural numbers]%
When working with well-scoped terms, we need to prove to the type checker that
scopes line up, which involves reasoning about the equality of natural number
expressions. Thus far, we have wanted to record exactly what properties are
needed that do not follow directly from their definitions. Therefore, we have
avoided the use of a special purpose solver; instead the library includes the
two monoid axioms about natural number addition, which must be used explicitly.
Note in particular that we do not rely on commutativity of additions; this
requires being careful about the order of arguments to additions when extending
the scope, e.g., as in the definition of the \texttt{Pat} type
(Section~\ref{sec:patterns}). Being careful about this has the benefit that, in
our experience, using commutativity becomes a sign that binders were added to
the scope in the wrong order. Despite our usage of the type system being
``Hasochism''~\cite{lindley:hasochism}, we would describe our experience as more
pleasurable than painful. This is partly due to our light usage of dependent
types, and partly due to Haskell's improvement in that regard, notably thanks to
the recent addition of new features, including explicit type application and
quantified constraints.

\item[Type inference]%
This work includes many operations that are polymorphic over types of
kind \cd{Nat -> Type} (i.e., scope-indexed types).  However, type inference is
more challenging when working with scoped patterns. There we use associated
types~\cite{associated-types} to indicate the number of variables bound by the
pattern because the parameter already refers to the scope of expressions that
appear inside the pattern itself. Unfortunately, associated types interact
poorly with unification and type class resolution. Our example suite includes
several positive examples to demonstrate the appropriate annotations needed to
guide the Haskell type checker.

\item[Multiple binding sorts] Our library is scope-safe for a \emph{single}
scope and does not support multiple sorts of scopes well. This causes
difficulties for languages such as System F~\cite{girard:thesis}, that bind both
type and term variables. While it is possible to index expressions by two
different scopes, using the library requires conversions to make sure that the
``right scope'' is the last parameter at certain times. Alternatively, users can
combine both type and term binding in the same scope.\  \rebound{} includes
examples of both approaches.

\item[Generic programming] \rebound{} uses \cd{GHC.Generics} to automatically
derive substitution and other operations. However, scope-safety causes two
complications, which our example \cd{PatGen.hs} demonstrates how to resolve.
First, generic programming is only available for datatypes that do not include
any ``existential'' variables. By isolating existentials into separate small
datatypes, users can provide these instances by hand while still retaining the
benefits of generic programming for the rest of their data structure. Second,
all type constructors used in the definition of the syntax must include their
scope as their last type argument.\@ \rebound{} provides an alternative
definition of the \cd{list} type to represent sequences of scoped expressions as
a scoped type.

\end{description}

\section{Related Work}%
\label{sec:related-work}

There are several binding libraries available for Haskell. The most similar to
this work is \Bound~\cite{kmett:bound}, which represents lambda calculus terms
using well-scoped de Bruijn indices~\cite{bird:debruijn}. In contrast to
\rebound, which makes extensive use of Dependent Haskell features, users of
\Bound{} represent syntax using a nested datatype and use a (derived) monad
instance for the type as the (single) substitution operation. \Bound{} adds
additional support for statically tracking scopes and optimizing the
implementation by delayed shifting. Examples distributed with the library show
that it can be used with debug names and in a language with pattern matching.

The \unbound{} library~\cite{weirich:binders,unbound-generics} uses a
\emph{locally nameless} representation for variable
binding~\cite{mcbride:locally-nameless}. In this approach, bound variables are
represented by de Bruijn indices and free variables are represented by names
(i.e., strings). When entering a new scope, users must replace bound variables
with fresh free variables. It also develops a library of pattern types to assist
in the definition of pattern binding. It does not track the scopes of free
variables statically.  The modern implementation of the
library~\cite{unbound-generics} uses generic
programming~\cite{magalhaes:generics} to automate the definition of these
operations and a freshness monad to generate free names.

The \emph{Foil} library~\cite{maclaurin:foil} uses phantom types to track the
scopes when variables are represented using \emph{names}. This use of names is
based on an algorithm called \emph{the rapier}~\cite{spj:rapier}, the approach used internally by
the GHC compiler. The key invariant is that names must be unique within their
scopes; on entering a new scope, binders must be renamed if they have already
been used. The \emph{Free Foil}~\cite{kudasov:free-foil} extension adds support
for pattern matching and uses TemplateHaskell~\cite{template-haskell} to
automate boilerplate definitions. To evaluate its expressiveness, Foil's authors
have used it to implement \texttt{lambda-pi}, a tutorial dependently-typed
language~\cite{lambda-pi}.

Several authors also describe how to implement binding structures in Haskell.
Augustsson's note~\cite{augustsson:cooked} provides simple Haskell
implementations using names, de Bruijn indices and higher-order abstract
syntax (HOAS). HOAS based embeddings can be modified using the type system to
rule out exotic terms~\cite{washburn:bgb-journal,kazutaka:embedding}, and
track scopes and types.  Bernardy and Pouillard~\cite{BernardyP13}
describe methodology for a scope-safe higher-order interface layered on top of
a de Bruijn indexed representation.

It is also possible to work with a well-\emph{typed} representation of syntax in
Haskell, which extends scope-safety with additional typing constraints for the
object language. Guillemette and Monnier demonstrates the encoding of a
type-preserving compiler~\cite{monnier:type-pres-compiler}.  Eisenberg's Stitch
functional pearl~\cite{eisenberg:stitch} includes a parser, type checker and
optimizer for a statically typed core language.  The Crucible
language~\cite{galois:crucible} uses a well-typed core language to build a suite
of static verifiers.

Several modern implementations of dependently-typed programming languages and
logics use de Bruijn indices to represent binding and make use of explicit
substitutions, including Idris 2, Agda, Twelf and Rocq. Idris 2 uses a
well-scoped abstract syntax type, with a simple representation of
substitutions as lists. In contrast, Agda, Twelf and Rocq optimize the
implementation of substitutions with explicit shifts, weakenings, and
liftings.

Going from language implementation to proofs, Cockx\footnote{
  \url{https://jesper.sikanda.be/posts/1001-syntax-representations.html}}
  provides an overview of variable representations that are possible in the
  dependently-typed language Agda.  While our approach based on well-scoped de
  Bruijn indices draws on similar work done in the context of a proof
  assistant~\cite{altenkirch-reus,debruijn-coq,allais:kit,erdi2018}, there are
  two important differences. First, type theories such as Agda, Rocq and Lean,
  require showing that substitution functions are total. This is most easily done by
  decomposing it into two steps: first renaming and then substitution, although there are
  techniques to combine these operations~\cite{altenkirch-reus}. In
  Haskell, we can define substitution in one go.  Second, when environments are
  delayed binders, $\alpha$-equivalent terms do not have a unique
  representation, depending on what environment is stored in the term.
  Therefore, additional care must be taken to make sure that all judgments are
  stable up to this equivalence.

\section{Conclusion}%
\label{sec:conclusion}

The \rebound{} library provides a framework for working with binding structures
that is, we believe, approachable to Haskell programmers. By statically tracking
scopes, \rebound{} eliminates many sources of confusion when working with de
Bruijn indices. Our library design isolates the complexity of binders using an
abstract type and automates the development of syntactic operations through
generic programming. This approach is expressive, as we have demonstrated by
using \rebound{} to implement languages with many different binding structures,
and by implementing several different operations using this representation.
These well-documented examples are part of the \rebound{} repository, and form
an extensive tutorial on working with well-scoped representations using
Dependent Haskell. Accompanying the examples is an extensive case study that
demonstrates an end-to-end use of the library in a practical setting. Finally,
we have evaluated the efficiency of \rebound{} against competing approaches and
have found that it outperforms its competitors, sometimes significantly.

Since the conducted evaluations yielded encouraging results in terms of both
time and space, we believe that additional case studies covering term
elaboration and higher-order unification should be performed. If these are
conclusive, we believe that the next evaluation step should then be to test the
library in an industrial-strength system.

There are a number of other potential avenues for future work. In particular, it
would be good to compare our environment representation more directly with the
approaches taken by Agda, Twelf, Rocq, and others. We would also like to explore
methodologies to improve Haskell's error reporting when code involving indexed
types, such as \rebound{}, fails to type check. Finally, we would like to
explore the use of a type-checker plug-in to automate reasoning about natural
number scopes.

\begin{acks}
Thanks to the members of IFIP WG 2.8 and the Haskell Symposium 2025 reviewers for comments and suggestions. 
Thanks to Francis Rinaldi for their feedback on this manuscript.
This material is based upon work supported by the
\grantsponsor{GS100000001}{National Science
Foundation}{http://dx.doi.org/10.13039/100000001} under Grant
No.~\grantnum{GS100000001}{NSF CCF-2327738}.  Any opinions, findings, and conclusions or
recommendations expressed in this material are those of the authors and do not
necessarily reflect the views of the National Science Foundation.
\end{acks}

\bibliographystyle{ACM-Reference-Format}
\bibliography{references}

\ifauthor{}
\pagebreak
\appendix

\section{Normalization Benchmark}%
\label{app:lennartb}

Normalization and evaluation benchmark, adapted from
Augusstson~\cite{augustsson:cooked}.
It calculates whether the Church encoding of $6!$
(i.e., 720) is equal to \cd{sum[0 .. 37] + 17}.
\begin{lstlisting}
let Zero = \z.\s.z;
    Succ = \n.\z.\s.s n;
    one = Succ Zero;
    two = Succ one;
    three = Succ two;
    isZero = \n.n true (\m.false);
    const = \x.\y.x;
    Pair = \a.\b.\p.p a b;
    fst = \ab.ab (\a.\b.a);
    snd = \ab.ab (\a.\b.b);
    fix = \ g. (\ x. g (x x)) (\ x. g (x x));
    add = fix (\radd.\x.\y.
                 x y (\ n. Succ (radd n y)));
    mul = fix (\rmul.\x.\y.
                 x Zero (\ n. add y (rmul n y)));
    fac = fix (\rfac.\x. x one (\ n. mul x (rfac n)));
    eqnat = fix (\reqnat.\x.\y.
                   x (y true (const false))
                       (\x1.y false (\y1.reqnat x1 y1)));
    sumto = fix (\rsumto.\x.
                   x Zero (\n.add x (rsumto n)));
    n5 = add two three;
    n6 = add three three;
    n17 = add n6 (add n6 n5);
    n37 = Succ (mul n6 n6);
    n703 = sumto n37;
    n720 = fac n6
in  eqnat n720 (add n703 n17)
\end{lstlisting}

\section{Normalization Algorithm}%
\label{app:nf}

Implementation of \cd{whnf} and \cd{nf} functions from \ref{sec:nf-benches}.
\begin{lstlisting}
-- | compute the weak-head normal form of open terms
whnf :: Exp n -> Exp n
whnf e@(Var _) = e
whnf e@(Lam _) = e
whnf (App f a) =
  case whnf f of
    Lam b -> whnf (instantiate b a) -- beta-reduction
    f' -> App f' a
\end{lstlisting}

\begin{lstlisting}
-- calculate the normal form of a term
nf :: Exp n -> Exp n
nf e@(Var x) = e
nf (Lam b)   = Lam (bind (nf (unbind b)))
nf (App f a) = case whnf f of
     Lam b -> nf (instantiate b a)  -- beta-reduction
     f' -> App (nf f') (nf a)
\end{lstlisting}
\fi

\end{document}